\documentclass[aps,preprint,nofootinbib]{revtex4}%
\usepackage{amsfonts}
\usepackage{amsmath}
\usepackage{amssymb}
\usepackage{lipsum}
\usepackage{xcolor}
\usepackage{amssymb}
\usepackage{amsmath}
\usepackage{slashed}
\usepackage{float}
\usepackage{physics}
\usepackage{graphicx}
\usepackage{hyperref}
\usepackage{dsfont}

\hypersetup{colorlinks}
\usepackage{cancel}
\usepackage[utf8]{inputenc}
\usepackage{stackrel}

\usepackage{float}
\usepackage{graphicx}%
\providecommand{\U}[1]{\protect\rule{.1in}{.1in}}

\begin{document}
 
\title{Exact flavored black $p$-branes and self-gravitating instantons from toroidal black holes with Skyrme hair}
\author{Patrick Concha$^{1}$, Carla Henr\'iquez-Baez$^{1,2}$, Evelyn Rodr\'iguez$^{1}$, Aldo Vera$^{2}$ \\
{\small $^{1}$\textit{Departamento de Matem\'atica y F\'isica Aplicadas,
Universidad Cat\'olica de la Sant\'isima Concepci\'on,
Alonso de Ribera 2850, Concepci\'on, Chile,}}\\
{\small $^{2}$\textit{Instituto de Ciencias F\'isicas y Matem\'aticas, Universidad
Austral de Chile, Casilla 567, Valdivia, Chile.}}\\
{\footnotesize patrick.concha@ucsc.cl, carla.henriquez@uach.cl, erodriguez@ucsc.cl, aldo.vera@uach.cl}}

\begin{abstract}
In this paper, using the maximal embedding of $SU(2)$ into $SU(N)$ in the Euler angles parameterization,  we construct a novel family of exact solutions of the Einstein $SU(N)$-Skyrme model. First, we present a hairy toroidal black hole in $D=4$ dimensions. This solution is asymptotically locally anti-de Sitter and is characterized by discrete hair parameters. Then, we perform a dimensional extension of the black hole to obtain black $p$-branes as solutions of the Einstein $SU(N)$-Non-linear sigma model in $D\geq5$ dimensions. These are homogeneous and topologically protected. Finally we show that, through a Wick rotation of the toroidal black hole, one can construct an exact self-gravitating instanton. The role that the flavor number $N$ plays in the geometry and thermodynamics of these configurations is also discussed. 
\end{abstract}

\maketitle


\section{Introduction}


The Skyrme model \cite{Skyrme1}, \cite{Skyrme2} is one of the most important effective non-linear field theories, since it is able to describe baryons in the low energy limit of Quantum Chromodynamics, where the usual perturbative methods can not be applied. In this model, baryons emerge as topological solitons (called Skyrmions) from the non-linear interactions between mesons, being the topological charge in this model the baryonic number \cite{Witten}, \cite{ANW}.
This is achieved from a purely bosonic action with isospin symmetry constructed from a scalar field $U(x)\in SU(N)$, that generalizes the Non-linear sigma model (NLSM).

When the Skyrme model is coupled to general relativity, it allows describing (self-)gravitating configurations supported by baryonic or mesonic matter (depending on whether the solutions are topologically trivial or not) such as black holes, gravitating solitons, and compact objects (Skyrme stars), among others. A relevant fact in this context is that the first counterexample to the well-known black hole no-hair conjecture was found in the Einstein $SU(2)$-Skyrme model \cite{Luckock:1986tr}, \cite{Droz:1991cx}, where spherically symmetric black hole solutions with non-trivial Skyrme fields were constructed numerically.\footnote{This black hole also turns out to be stable under spherical linear perturbations \cite{Heusler:1992av}.}

In general, constructing solutions in the Skyrme model (and, of course, in the Einstein-Skyrme model) is a complicated task due to the high non-linearity of the field equations, together with the fact that the Skyrme-BPS bound cannot be saturated in generic cases.\footnote{This is different from what happens in the Yang-Mills theory, where the saturation of the BPS bound has allowed the construction of several solutions (see \cite{Manton:1981mp}, \cite{MantonBook} and \cite{OliveBook}).} In consequence, most of the known solutions have been found using numerical methods (see \cite{Bizon}, \cite{Sawado}, \cite{Kunz}, \cite{BD}, \cite{Brihaye}, \cite{Nelmes} and references therein). However, in recent years, the development of new techniques has allowed a good number of analytical solutions, both in flat space-time and in the (self-)gravitating case. Within these, we can highlight the construction of crystals of (superconducting) baryonic tubes and layers at finite volume \cite{Pedro}, \cite{crystals}, \cite{Diego}, pionic black holes with different geometries of the event horizon \cite{CanforaMaeda}, \cite{toroidal1} (see also \cite{Eiroa}, \cite{Daniel} and \cite{accretion}), self-gravitating Skyrmions \cite{Eloy1}, \cite{Eloy2}, \cite{gravtube1}, \cite{gravtube2}, and extended black objects such as black strings \cite{toroidal2}, \cite{toroidal3}, \cite{Giacomini:2019qov}. 

It is important to note that almost all the solutions mentioned above (both analytical and numerical) have been constructed for the internal symmetry group $SU(2)$, that is, the two flavors case. Finding solutions in the Skyrme model for $N>2$ has an important extra complexity (as can be seen in Refs. \cite{Eloy2}, \cite{Bala1}, \cite{Bala2}, \cite{Ioa}), since in principle, it requires to solve a set of $(N^2-1)$ coupled non-linear field equations.

Here we are mainly interested in the construction of analytical black objects for arbitrary values of the flavor number $N$ -in particular, toroidal black holes and black $p$-branes- and in the analysis of their geometric and thermal properties. For this purpose, a good approach is to consider some relatively simple embedding of $SU(N)$ to construct the matter field $U$, but that at the same time, it will not be a trivial embedding of $SU(2)$ into $SU(N)$. In Refs. \cite{euler1}, \cite{euler2}, \cite{euler3} an ansatz was introduced that fulfilled these requirements; this is the maximal embedding of
$SU(2)$ into $SU(N)$ in the Euler angles parametrization. Recently, this maximal embedding has shown to be very useful in the construction of analytical solutions both in the Skyrme model and in the Yang-Mills theory  \cite{toroidal3}, \cite{Gomberoff}, \cite{euler4}, \cite{SU(N)}, so it will be our starting point in this work.

 In particular, black holes with flat horizons and negative cosmological constant are an interesting class of solutions which have attracted a lot of attention due to their applications in holography \cite{Papantonopoulos:2011zz}, \cite{Caldarelli:2016nni}. This is due to the possibility of using the AdS/CFT correspondence to describe interesting field theories on the boundary of the black hole space-time. For instance, the quark-gluon plasma (QGP) is modeled, via holography, by a field theory dual to a thermal anti-de Sitter black hole. Since the QGP exists in Minkowski space, one needs to use black holes with topologically planar event horizons. 
On the other hand, for black brane solutions \cite{Horowitz:1991cd} the thermodynamics can be extended to hydrodynamics. Indeed, the hydrodynamic regime is meaningful only for translationally-invariant horizon. From the holographic principle point of view, a black brane corresponds to a certain finite-temperature quantum field theory in fewer numbers of space-time dimensions, and the hydrodynamic behavior of a black brane horizon is identified with the hydrodynamic behavior of the dual theory (see \cite{Son:2007vk} for a review about the connection, via the AdS/CFT correspondence, between hydrodynamics and black hole physics). Then, the hairy black holes and black branes solutions which will be constructed in this work could have many interesting applications in the holography context.

The paper is organized as follows: In Sec. \ref{sec-2} we give a brief summary of the Einstein $SU(N)$-Skyrme model, and we introduce our general ansatz for the Skyrme field. In Sec. \ref{sec-3} we construct analytic hairy toroidal black holes, and we study the thermodynamics of the solution, emphasizing the role played by the flavor number. In Sec. \ref{sec-4} we construct black $p$-branes and a self-gravitating instanton solutions using the toroidal black hole as a starting point. Section \ref{sec-5} is dedicated to the conclusions.


\section{Preliminaries} \label{sec-2}


\subsection{The Einstein $SU(N)$-Skyrme model} 

The Einstein $SU(N)$-Skyrme model in $D=4$ dimensions is described by the following action 
\begin{equation}
I[g,U]=\int d^{4}x\sqrt{-g}\left( \frac{R-2\Lambda }{2\kappa }+%
\frac{K}{4}\mathrm{Tr}[L^{\mu }L_{\mu }]+\frac{K\lambda }{32}\mathrm{Tr}%
\left( G_{\mu \nu }G^{\mu \nu }\right) \right) \ .  \label{I}
\end{equation}%
The first part in Eq. \eqref{I} is the gravitational section, where $R$ is the Ricci scalar, $g$ is the determinant of the metric $g_{\mu\nu}$, $\kappa$ is the gravitational constant, and $\Lambda$ is the cosmological constant. The second part of the action corresponds to the Skyrme action, where $K$ and $\lambda$ are positive coupling  constants that are experimentally determined.\footnote{The case $\lambda \rightarrow 0$ corresponds to the Einstein $SU(N)$-NLSM theory, which will be relevant in Sec. \ref{sec-4} in the construction of higher dimensional solutions.} 
Here $L_{\mu}$ are the Maurer-Cartan left-invariant form components, given by 
\begin{equation}
    L_{\mu}=U^{-1}\partial_{\mu} U = L^{i}_{\mu}t_{i} \ , \label{Maurer-Cartan-form}
\end{equation}
where $U\left(x \right) \in SU(N)$, where $N$ is the flavor number. The matrices $t_{i}$ are the generators of the $SU(N)$ Lie group, with $i=1, \ldots , \left( N^2-1 \right)$, and the $G_{\mu\nu}$ tensor is defined as 
\begin{equation}
G_{\mu \nu }=\left[ L_{\mu },L_{\nu }\right] \ .
\end{equation}
The energy-momentum tensor of the system is 
\begin{equation}
T_{\mu \nu }=-\frac{K}{2}\mathrm{Tr}\left[ L_{\mu }L_{\nu }-\frac{1}{2}%
g_{\mu \nu }L^{\alpha }L_{\alpha }\right. \,+\left. \frac{\lambda }{4}\left(
g^{\alpha \beta }G_{\mu \alpha }G_{\nu \beta }-\frac{g_{\mu \nu }}{4}%
G_{\sigma \rho }G^{\sigma \rho }\right) \right] \ .  \notag  \label{timunu1}
\end{equation}%
The Einstein $SU(N)$-Skyrme field equations are obtained varying the action in Eq. \eqref{I} with respect to the fundamental fields $g_{\mu\nu}$ and $U$, obtaining
\begin{gather}
\nabla _{\mu }L^{\mu }+\frac{\lambda }{4}\nabla _{\mu }[L_{\nu },G^{\mu \nu
}]=0\ , \\
R_{\mu \nu }-\frac{1}{2}Rg_{\mu \nu }+\Lambda g_{\mu \nu }=\kappa T_{\mu \nu
}\ .
\end{gather}
In the Skyrme theory, relevant properties of the solutions  are encoded in the topological charge, which is defined as
\begin{equation}
B = \frac{1}{24\pi^2} \int  \rho_B \ , \quad \rho_B=\varepsilon _{ijk} \text{Tr}\left\{ \left(
U^{-1}\partial ^{i}U\right) \left( U^{-1}\partial ^{j}U\right) \left(
U^{-1}\partial ^{k}U\right) \right\} \ ,  \label{B}
\end{equation}%
where $\{i,j,k \}$ are spatial indices.

\subsection{Skyrme field in the Euler parametrization}

In order to construct analytical solutions of the Einstein $SU(N)$-NLSM and Skyrme model, we will use for the matter field $U$ the so-called maximal embedding ansatz \cite{euler1}, \cite{euler2}, \cite{euler3}. As is well-known, there are many ways of embedding $SU(2)$ into $SU(N)$, but not all of these are fruitful to build exact solutions. Here we choose the "maximal one," which gives rise to an irreducible spin-$j$ representation of $SU(2)$ of spin $j = (N - 1) / 2$. 

The matter field $U(x)\in SU(N)$ in the generalized Euler angles parametrization reads  
\begin{equation}
    U=e^{F_{1}\left(x^{\mu} \right)\cdot T_{3}}e^{F_{2}\left(x^{\mu}\right) \cdot T_{2}} e^{F_{3}\left(x^{\mu}\right) \cdot T_{3}} \ ,  \label{matter-ansatz}
\end{equation}
where $T_{i}$ (with $i=1,2,3$) are three matrices of a representation of the Lie algebra $\mathfrak{su}(N)$. The generators of this $three$-dimensional subalgebra of $\mathfrak{su}(N)$ satisfy the following relations
\begin{eqnarray*}
\left[T_{j}, T_{k} \right]&=& \epsilon_{jkm}T_{m} \ , \\
\text{Tr}\left(T_{j}T_{k} \right)&=&-\frac{N\left(N^2-1 \right)}{12}\delta_{jk} \ ,
\end{eqnarray*}
and they are explicitly given by 
\begin{align}
T_1&=-\frac{i}{2}\sum_{j=2}^{N} \sqrt{(j-1)(N-j+1)}(E_{j-1,j}+E_{j,j-1}) \ ,
\label{T1} \\
T_2&=\frac{1}{2}\sum_{j=2}^{N} \sqrt{(j-1)(N-j+1)}(E_{j-1,j}-E_{j,j-1}) \ ,
\label{T2} \\
T_3&=i\sum_{j=1}^{N} (\frac{N+1}{2}-j)E_{j,j} \ ,  \label{T3}
\end{align}
with $(E_{i,j})_{mn}=\delta_{im}\delta_{jn}$\, where $\delta_{ij}$ is the Kronecker delta.\footnote{Note that the $T_i$ matrices are anti-hermitian. But, one can easily recover a hermitian set by multiplying the matrices by $i$. Of course, in order to obtain the same solutions presented below we need to multiply the $F_i$ function also by $i$.} 
As the above matrices conform an irreducible representation of $SU(2)$, the solutions constructed in the next sections using this formalism will be non-trivial embedding of $SU(2)$ into $SU(N)$, and therefore the role of the flavor number $N$ can be explicitly showed.\footnote{The key point in this construction is that, actually, the map between the Lie groups is not an embedding of $SU(2)$ into $SU(N)$, but just of $S_3$ into $SU(N)$ (see \cite{euler1}, \cite{euler2}, \cite{euler3}).} 

One way to see that the solutions constructed using this formalism are not simply solutions for the usual case of two flavors, is by noting that
\begin{equation}
 (\vec{T})^2= \rho(N) \mathds{1} \ , \qquad \rho(N)=-\frac{N^2-1}{4} \ , \label{Tcua}
\end{equation}
with $\mathds{1}$ the $N\times N$ identity matrix. Then, choosing the irreducible representation of $SU(2)$ for all values of $N$ implies, that for every $N$, we are using a representation with different spin. As a consequence of Eq. \eqref{Tcua}, it is expected that the trace in the Skyrme action leads to the relevant physical quantities of the solutions depending explicitly on $N$. In fact, it is useful to define the following quantity
\begin{equation}  \label{aN}
    a_{N}=\frac{N\left(N^2 -1 \right)}{6} \ ,
\end{equation}
which comes from the calculation of the trace of the generators that will appear in all the solutions presented below. We will see later that both the geometry and the thermodynamics of the solutions depend strongly on $N$. It is important to point out that this construction has recently been used to find both gravitating and flat space-time solutions in non-Abelian theories
\cite{toroidal3}, \cite{Gomberoff}, \cite{euler4}, \cite{SU(N)} (see \cite{euler4} for a nice review on applications to nuclear physics of this formalism). The complete mathematical formulation of the maximal embedding in the Euler angles parametrization can be found in Refs. \cite{euler1}, \cite{euler2} and \cite{euler3}. 

On the other hand, the functions $F_{i}$ in Eq. \eqref{matter-ansatz} can depend on all the coordinates, but they are chosen such that they solve the Einstein $SU(N)$-Skyrme equations system. In particular we will demand that $F_2=F_2(\theta)$ and $F_3=F_3(\phi)$, where the angular coordinates $\{\theta$, $\phi\}$ have the ranges 
\begin{equation} \label{ranges}
 0\leq\theta < \pi \ , \quad  0\leq \phi < 2\pi \ .
\end{equation}
From the above, and replacing Eq. \eqref{matter-ansatz} in Eq. \eqref{B}, one can check that the topological charge density goes as 
\begin{equation}
    \rho_B \sim \biggl\{ \sin(F_2) \partial_\theta F_2 \partial_\phi F_3 F_1'  \biggl\} dF_1 \wedge d F_2 \wedge F_3  \ ,
\end{equation}
where the prime denotes derivative with respect to coordinates other than
$\theta$ and $\phi$. Here we can see that necessary (but not sufficient) conditions to have a non-zero topological charge are 
\begin{equation} \label{cond}
    dF_1 \wedge d F_2 \wedge d F_3 \neq 0 \ ,\quad  \text{and} \quad  F'_1 \neq 0 \ . 
\end{equation}


\section{Analytic toroidal black hole}  \label{sec-3}


In this section we construct an analytical solution describing a toroidal black hole with Skyrme hair and arbitrary $N$ in $D=4$ dimensions, and we show its main physical properties. We provide the thermal analysis of the solution and some comments about the role of the flavor number $N$, as most of the thermodynamic variables depend in a non-trivial way on it.

\subsection{Constructing a hairy toroidal black hole from the maximal embbeding}

From Refs. \cite{toroidal1}, \cite{toroidal2} and \cite{toroidal3} we know that a good ansatz for the matter field in Eq. \eqref{matter-ansatz} that allows to construct toroidal black holes is the following 
\begin{equation} \label{matter1}
 F_1(x^\mu) = 0 \ , \qquad F_2(x^\mu) =  q \theta \ , \qquad F_3(x^\mu) = v \phi \ ,   
\end{equation}
where $q$ and $v$ are integer numbers (according to the Euler parametrization; see \cite{euler1}, \cite{euler2}, \cite{euler3}) that we recognize as hair parameters. The space-time is described by a static metric with toroidal geometry 
\begin{equation}  \label{g1}
 ds^2 =  -f(r) dt^2+ \frac{1}{f(r)} dr^2 + r^2 d\theta^2 + c^2 r^2 d\phi^2 \ ,
\end{equation}
where $c$ is a constant to be found.
It is a direct computation to show that, with the above Ansatz, the Skyrme field equations are automatically satisfied for all values of $N$.
On the other hand, the Einstein equations are reduced to just one integrable equation that can be directly solved, obtaining
\begin{equation} \label{f1}
 f(r)= -b_1 -\frac{m}{r}  +\frac{b_2}{r^2} - \frac{\Lambda}{3} r^2 \ , 
\end{equation}
where $m$ is an integration constant, and the constants $b_1$ and $b_2$ are fixed -in the terms of the couplings and the hair parameters- as follows
\begin{equation}
 b_1= \frac{K \kappa q^2 a_N}{4} \ , \qquad b_2 = \frac{K \kappa q^4 \lambda a_N}{32} \ ,   \label{bs}
\end{equation}
where $a_N$ has been defined in Eq. \eqref{aN}. Also, the constant $c$ in Eq. \eqref{g1} turns out to be $c^2= \frac{v^2}{q^2}$ according to the Einstein equations.\footnote{Note that a conical singularity can exist in the metric due to the presence of the constant $c^2$ in front of the torus metric in Eq. \eqref{g1}. Now, this conical singularity can be removed from the metric considering that the parameters $v$ and $q$ are equal. This leads to the elimination of one of the hair parameters in the solution, but not both at the same time. In fact, from Eqs. \eqref{f1} and \eqref{bs}, it can be seen that the function $f$ that characterizes the toroidal black hole solution depends on only one of these parameters, which therefore cannot be removed from the solution by making $v=q$. One can lead to the same conclusion by trying to redefine the $\phi$ coordinate.}

This configuration represents an asymptotically locally (Anti-)de Sitter toroidal black hole with Skyrme hair allowing arbitrary values of the flavor number (as can be seen from Eq. \eqref{bs}, where $N$ appears explicitly through $a_N$). 
This analytic solution is the generalization of the toroidal black holes reported in Ref. \cite{toroidal1} (see also \cite{toroidal2} and \cite{toroidal3}), that includes the Skyrme term as well as an arbitrary number of flavors. 
Although the radius of the horizon can be computed analytically from Eq. \eqref{f1}, the expression is very complicated and is not needed for our purposes. The case with $\lambda=0$ is particularly simple, and can be seen in detail in Ref. \cite{toroidal3}.

It is important to note that this black hole solution has zero topological charge because it does not satisfy the requirements in Eq. \eqref{cond}.
In the following section we will show that black $p$-branes constructed from the toroidal black hole presented above can satisfy this condition, and lead to topologically non-trivial solutions of the Einstein $SU(N)$-NLSM theory in $D\geq5$ dimensions.

\subsection{Thermodynamics}

Here we perform the thermal analysis of the toroidal black hole constructed above using the Euclidean approach \cite{Gibbons:1977mu} . The formulation is straightforward; upon replacing $t\rightarrow i\tau$ in the metric in Eq. \eqref{g1} we obtain the Euclidean black hole metric. The temperature is the inverse of the  Euclidean time period, which turns out to be
\begin{equation}\label{Temp1}
    T\equiv\beta^{-1}=\frac{1}{4\pi}f^{\prime}\left(r_{+}\right)=-\frac{K\kappa a_{N}q^{2}}{16 \pi r_{+} }-\frac{K\kappa a_{N} \lambda q^{4}}{128 \pi r_{+}^{3}}-\frac{\Lambda r_{+}}{4 \pi} \ ,
\end{equation}
where $r_{+}$ is the largest root of the equation $f\left(r_{+}\right)=0$. From the Euclidean action (see Appendix \ref{App1}) we can obtain the free energy
\begin{equation}
     F=\beta^{-1}I^{\text{E}}=-\frac{ v\pi r_{+}}{32q}\left(K\kappa\alpha_{N}q^{2}-\frac{4}{3}\Lambda r_{+}^{2}-\frac{3K\kappa \alpha_{N}\lambda q^{4}}{8 r_{+}^{2}}\right)\,.\label{fe}
\end{equation}
In the grand canonical ensemble, the thermodynamic quantities satisfy
\begin{equation}
    F=E-TS\,,
\end{equation}
where $E$ is the energy and $S$ is the entropy. In particular, the thermodynamic variables of the system satisfy
\begin{equation*}
      E=\frac{\partial I^{\text{E}}}{\partial\beta} \ ,  \qquad S=\beta \frac{\partial I^{E}}{\partial\beta}-I^{\text{E}} \ , 
\end{equation*}
where $I^{\text{E}}$ is the Euclidean action given in Eq. \eqref{Euclact}. Considering the above relations, we can show that the thermodynamics variables are given as follows,\footnote{Note that the energy computed from the Euclidean action matches the one computed using the holographic method (see Appendix \ref{App2}).}
 \begin{align}
      E&=-\frac{v\pi r_{+}}{48q}\left(3K\kappa a_{N}q^{2}+4\Lambda r_{+}^{2}-\frac{3K\kappa \alpha_{N}\lambda q^{4}}{8 r_{+}^{2}}\right)\,, \label{mass}\\
      S&=\frac{\pi^{2}v\, r_{+}^{2}}{2q}=\frac{A_{h}}{4}\,, \label{ent}
\end{align}
where $A_{h}$ denotes the area of the event horizon. Note that the entropy does not depend on the flavor number, but it depends on $v$ and $q$. Therefore, phase transitions are expected to occur varying the values of the hair parameters. 

 As it is expected, using Eq. \eqref{Temp1} and defining the mass as $M=E$,  the first law of black hole thermodynamics is satisfied, i.e.,
\begin{equation}
    \delta M=T\delta S \ , 
\end{equation}
where the cosmological constant $\Lambda$ is treated as a fixed parameter.
It is well-known that a general study of the classical stability of hairy black holes is not straightforward to approach. Here we will analyze the local thermodynamic stability of the above solution with respect to thermal fluctuations by computing the heat capacity and looking at if it is positive. For the toroidal black hole with Skyrme hair we found
\begin{equation}
    C=T\left(\frac{\partial S}{\partial T}\right)=\frac{\pi^{2} v r_{+}^{2}}{q}\left(\frac{32 \Lambda r_{+}^{4}+8K\kappa a_{N}q^{2}r_{+}^{2}+K\kappa a_{N}\lambda q^{4}}{32 \Lambda r_{+}^{4}-8K\kappa a_{N}q^{2}r_{+}^{2}-3K\kappa a_{N}\lambda q^{4}}\right)\,.
\end{equation}
Then, in order to have a positive $C$, the following constraint must be satisfied
\begin{equation}
    r_{+}>\frac{\sqrt{a_{N}K \kappa q^{2}+\sqrt{a_{N}^{2}K^{2}\kappa^{2}q^{4}-2a_{N}K \kappa \lambda \Lambda q^{4}}}}{2 \sqrt{2}\sqrt{-\Lambda}}\,.\label{rmascal}
\end{equation}
We can see that the above constraint is always satisfied if the cosmological constant $\Lambda$ takes negatives values, so that the asymptotically locally anti-de Sitter black hole can be thermally stable. In addition, as we will see bellow, Eq. \eqref{rmascal} is always satisfied and the heat capacity is always positive when we demand to have a real solution to the equation $f\left(r_{+}\right)=0$. Naturally, all the thermodynamic quantities previously defined, reduce to those found in Ref. \cite{toroidal3} when we set $\lambda=0$. 

\subsection{On the role of $N$}

\begin{figure}[h!]
    \centering
    \includegraphics[scale=1]{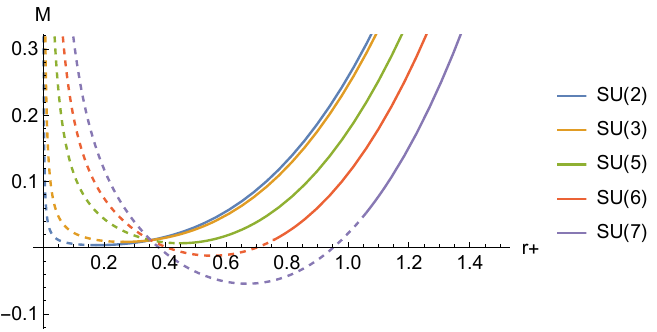} 
    \caption{Mass $M$ of the toroidal black hole with Skyrme hair as a function of the event horizon for different values of the flavor number. The dashed lines highlight the mass evaluated at event horizon radii which are smaller than the minimum value it can take in order to have real solutions.}
    \label{fig1}
\end{figure}

In the previous subsection it has been seen that both the thermal and geometric properties of the black hole strongly depend on the number of flavors in the theory. For instance, according to Eq. \eqref{mass}, the mass of the black hole depends in a non-trivial way on $N$. We show this dependence in Fig. \ref{fig1}, where we have plotted the mass as a function of the event horizon of the black hole for different values of $N$. In the analysis we have set $\lambda=1$, $K=1$, $\Lambda=-1$ and $\kappa=1/40$. Moreover, for simplicity, we set $q=v=1$.
From Fig. \ref{fig1} (or looking at Eq. \eqref{mass}) we can see that $M$ is always positive for $N\leq5$, while for $N\geq6$ there are sectors where the mass can take negative values. However, if we consider the extremal value $r_{\text{min}}=r_{\text{min}}(N)$ allowing for a real solution to the equation $f(r_{+})=0$, these negative values of the mass are not permitted. This is depicted with dashed lines in Fig. \ref{fig1}.  
We can also see that, for small radii of the event horizon, the most massive configurations are those with the largest number of flavors. This behavior reverses after a critical point (which does not depend on $N$) for large event horizon radii.
\begin{figure}[h!]
    \centering
    \includegraphics[scale=1]{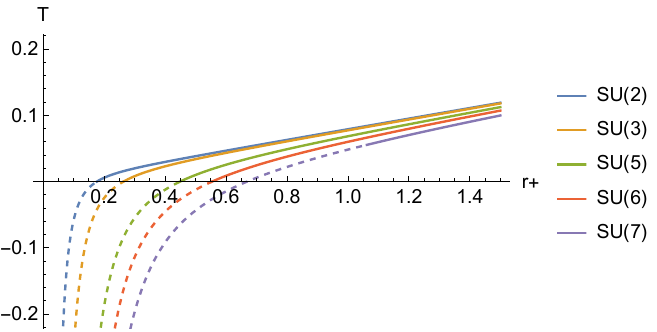} 
    \caption{Temperature $T$ of the toroidal black hole with Skyrme hair as a function of its event horizon for different values of $N$. At a certain horizon radius value depending on $N$, the temperature becomes positive. The dashed lines indicate the values of $T$ for $r_{+}<r_{\text{min}}$.  For a fixed $r_{+}$, $T$ decreases as the value of $N$ increases.}
    \label{fig2}
\end{figure}
As we have mentioned before, the entropy $S$ does not depend explicitly on the value of $N$, as we can see from Eq. \eqref{ent}. Furthermore, its behavior is not affected by the Skyrme term. On the other hand, the temperature of the solution as a function of the event horizon for different values of $N$ is plotted in Fig. \ref{fig2}. For a fixed value of $r_{+}$, we have that $T$ decreases as we increase $N$. Regarding the free energy computed in Eq. \eqref{fe}, it is possible to find an analytical (but complicated) expression for it in terms of the temperature. However, as it is not needed for our purposes, we instead provide it with the graph $F(T)$ for different values of $N$ (see Fig. \ref{fig3}). From the plot we can see that as we increase the temperature, transitions start to happen for $N\leq5$. For $N\geq6$ there is no change of sign in the free energy, independent of the value of the temperature. Unlike the particular case $\lambda=0$ studied in Ref. \cite{toroidal2}, here we can see that the presence of the Skyrme term allows us to have transitions, at least for some values of $N$. Indeed, for the flat black hole constructed in \cite{toroidal2}, there are no transitions and the favored configuration is always the one with the higher flavor number. We can also see that for high values of temperature the free energy is lower as we increase $N$.
\begin{figure}[h!]
    \centering
    \includegraphics[scale=1]{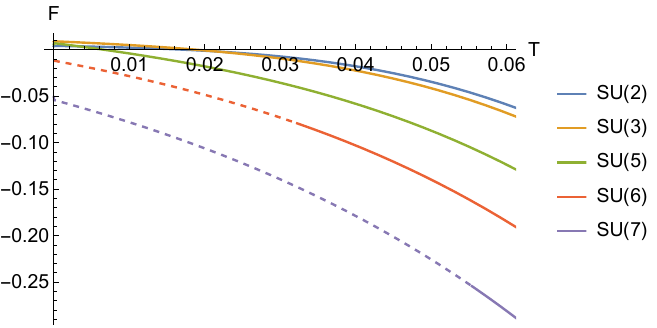} 
    \caption{Free energy $F$ of the toroidal black hole with Skyrme hair as a function of temperature for different values of $N$. As we increase the temperature, transitions start to take place for $N\leq5$, and for high enough values of $T$ the free energy is lower as $N$ increases. The dashed lines represent the sectors which are not allowed under the requirement of having a real solution. }
    \label{fig3}
\end{figure}

\begin{figure}[h!]
    \centering
    \includegraphics[scale=1]{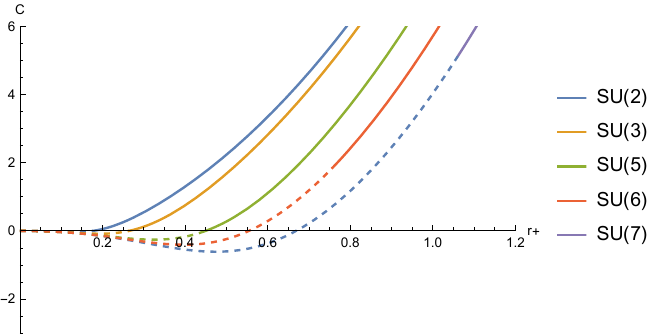} 
    \caption{Heat capacity $C$ of the toroidal black hole with Skyrme hair as a function of the event horizon for different values of $N$.}
    \label{fig4}
\end{figure}

From Fig. {\ref{fig4}} we see that the heat capacity for the toroidal black hole solution is always positive for the values allowing a real solution.  Then, we can say that the toroidal black hole solution is thermodynamically stable.

The analytic hairy toroidal black holes previously constructed could have interesting applications in the context of holography. Indeed, black holes with flat horizon have received great attention due to the celebrated AdS/CFT conjecture, and the possibility of using it in the description of field theories on the boundary of the black hole space-time itself. 
In the next section, we extend our study to the construction of homogeneous black $p$-branes and self-gravitating instantons. For the former, we also include the thermodynamic analysis. In particular, we provide with the thermodynamic quantities and some comments about stability are mentioned.


\section{Black $p$-branes and self-gravitating instantons from toroidal black holes} \label{sec-4}


In this section we will show that solutions describing homogeneous black $p$-branes with non-trivial topological fluxes as well as self-gravitating instantons, can be  constructed in a direct way from the toroidal black hole. In particular, black $p$-branes are obtained through a dimensional extension of the metric of the black hole adding extended homogeneous directions and fields living on those extra dimensions with a linear dependence on its coordinates, while self-gravitating instantons emerge through a Wick-like rotation of the space-time coordinates.

\subsection{Homogeneous black $p$-branes in the Einstein $SU(N)$-NLSM}

Homogeneous black $p$-branes in the Einstein $SU(N)$-NLSM model in $D\geq5$ space-time dimensions can be constructed through a dimensional extension of the toroidal black hole presented above, described by the metric in Eq. \eqref{f1}, as follows. For the matter field we consider the following ansatz
\begin{equation}  \label{Ubs}
 F_1(x^\mu) = \sum_{i=1}^{p} c_i z^i \ , \qquad F_2(x^\mu) =  q \theta \ , \qquad F_3(x^\mu) = v \phi   \ ,   
\end{equation}
where $c_i$ are hair parameters and $z^{i}$ correspond to the coordinates of the $p$-extended directions. On the other hand, let us consider the following space-time metric in $D=(4+p)$ dimensions 
\begin{equation} \label{pbrane}
    ds^2 = -f(r) dt^2 + \frac{1}{f(r)} dr^2 + r^2 d\theta^2 + \frac{v^2}{q^2}r^2 d\phi ^2 + H_{p}\sum_{i=1}^{p}c_{i}^2 dz_i dz^i \ ,  
\end{equation}
which corresponds to the black hole ansatz in Eq. \eqref{g1} plus $p$ extended homogeneous directions parametrized by the $z_i$ coordinates, 
with $H_{p}$ being a constant defined as follows
\begin{equation}  \label{Hp}
    H_{p}= \frac{(p+2) K \kappa a_{N}}{8 (-\Lambda)}  \ . 
\end{equation}
With the above, one can check that the Einstein $SU(N)$-Skyrme field equations demand that 
\begin{equation}
    \lambda = 0 \ ,  
\end{equation}
so that the solution we are looking for belongs to the Einstein $SU(N)$-NLSM. In fact, considering the above constraint, the $D$-dimensional Einstein equations are solved by 
\begin{equation}
    f(r) =   -\frac{K\kappa q^{2}a_{N}}{4} +\frac{m}{r} - \frac{2 \Lambda}{3\left(p + 2\right)} r^2 \ ,
\end{equation}
with $m$ an integration constant.\footnote{This construction is similar to the one reported in Ref. \cite{CisternaOliva} where the authors construct homogeneous black string in general relativity with negative cosmological constant by considering massless scalar fields that are linear in the extended directions \cite{Cisterna:2019scr}. See also \cite{Cisterna:2018mww} and \cite{Cisterna:2020kde} for the extension to Einstein-Gauss-Bonnet and Lovelock theories.} From Eq. \eqref{Hp} is clear that for these solutions the cosmological constant must take negative values, therefore, again we have asymptotically anti-de Sitter configurations. 
The event horizon of the solution is localized at 
\begin{small}
\begin{equation}
r_{+}= \frac{a_{N} \kappa  K \Lambda  (p+2) q^2-\left(\sqrt{\Lambda ^3 (p+2)^2 \left(a_{N}^3 \kappa ^3 K^3 (p+2) q^6+144 \Lambda  m^2\right)}+12 \Lambda ^2 m (p+2)\right)^{2/3}}{2 \Lambda 
   \sqrt[3]{\sqrt{\Lambda ^3 (p+2)^2 \left(a_{N}^3 \kappa ^3 K^3 (p+2) q^6+144 \Lambda  m^2\right)}+12 \Lambda ^2 m (p+2)}} \ .
\end{equation}
\end{small}
From the previous expression we see that, in order to have a real square root, the integration constant $m$ must satisfy the following condition
\begin{equation}\label{rest:m-bs-real}
    m> \frac{q^3 \sqrt{(p+2)a_{N}^3 \kappa^3 K^3}}{\sqrt{288(-\Lambda)}} \ .
\end{equation}
It is important to mention that, even when the matter field depends explicitly on the extended coordinates $z_i$, the energy density for the solution is finite because it does not depend on the extra dimensions. 
Additionally, one of the main characteristics that the homogeneous black $p$-branes constructed here possess is that they are topologically non-trivial. In fact, the presence of the extra extended directions in Eq. \eqref{Ubs}, in addition to the two angular coordinates in the matter field, allows to integrate the topological charge density defined in Eq. \eqref{B} on a space-like hypersurface to obtain a non-zero value of $B$. A simple way to see this is by compactifying the extended directions of the branes. For simplicity, let us consider the case of the black string solution in $D=5$ dimensions (the case with $p=1$). As the conditions in Eq. \eqref{cond} are satisfied, one can integrate $\rho_B$ at fixed $t$ and fixed $r$ in the ranges in Eq. \eqref{ranges} for the angular coordinates, and considering the following range
\begin{equation*}
     0 \leq z < 4 \pi \ , 
\end{equation*}
for the extended direction. Then, integrating the topological charge density in Eq. \eqref{B} we obtain the topological charge
\begin{equation} \label{Bbrane}
    B= v c_{1} \ , 
\end{equation}
where we have demanded that $q$ is an odd number. Thus, Eq. \eqref{Bbrane} determines that the black string is topologically protected (and the resulting charge is an integer number).\footnote{However, this topological charge cannot be directly interpreted as the baryon number in the Skyrme model, since in this case the integration on $r$ must be performed. In this sense, $B$ in Eq. \eqref{Bbrane} must be thought as the value of the flux in the $r$-direction at constant time.} The generalization to the black $p$-branes case is direct. The existence of a topological charge is a very important point because its existence is usually related to stability, suggesting that the black $p$-branes constructed here could be stable. In fact, black string and black $p$-branes solutions turn out to be unstable under perturbations \cite{GL1}, \cite{GL2} (see also \cite{LehnerPretorius}, \cite{Suzuki:2015axa}, \cite{Emparan:2015gva} and references therein). Of course, classical stability of our solutions requires an exhaustive perturbative analysis. We hope to come back to this important issue in a future publication.

For black $p$-branes it is possible to calculate the thermodynamic quantities following the same steps as it was for the case of the black hole presented in the previous section. 

\begin{figure}[h!] 
    \centering
    \includegraphics[scale=0.9]{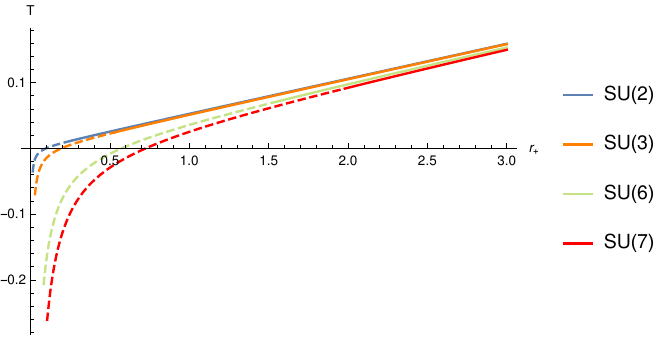} 
    \caption{Temperature $T$ of the black string, $p=1$,  as a function of the event horizon for different values of the flavor number. }
    \label{figbsT}
\end{figure}

The temperature for the black string $p$-branes reads
\begin{equation}\label{T-BS}
 T = -\frac{a_{N}K\kappa q^2}{16 \pi r_{+}}-\frac{r_{+}\Lambda}{2(p+2)\pi}  \ .
 \end{equation}
For this solution the temperature is always positive over the range of allowed values of the event horizon in order to have a real solution, which is given by restriction on Eq. \eqref{rest:m-bs-real}. Also, we see that the temperature $T$ decreases as the value of the flavor number increases; see Fig. \ref{figbsT}. In this section, for the plots, 
we have considered the simplest case in the space of solutions, that is, a black string (the $p=1$ case), and we have set the same values for the couplings constants and the hair parameters as in the black hole case. 

\begin{figure}[h!]
    \centering
    \includegraphics[scale=0.9]{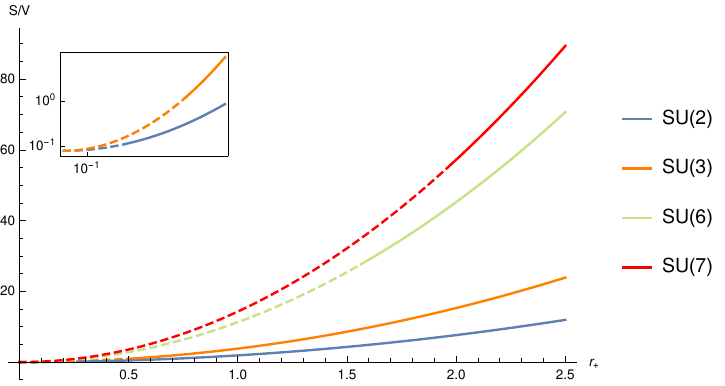} 
    \caption{Entropy density $s=\frac{S}{V_{p}}$ of the black string as a function of the event horizon for different values of the flavor number. The dotted lines represent the values of $S$ for $r_{+}$ outside the range allowed to have real solutions.}
    \label{figbsS}
\end{figure}

\begin{figure}[h!]
    \centering
    \includegraphics[scale=0.9]{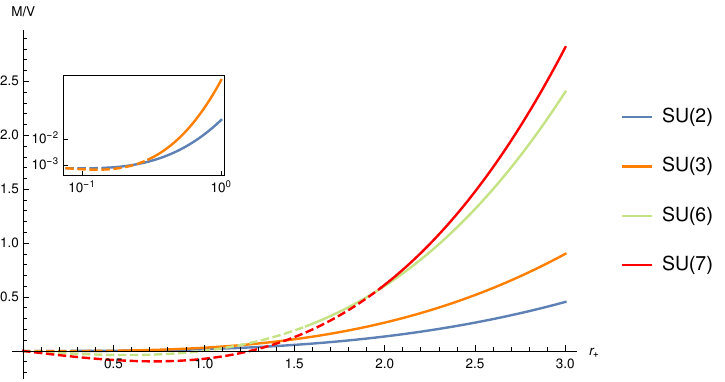} 
    \caption{Mass density $m=\frac{M}{V_{p}}$ of the black string as a function of the event horizon for different values of the flavor number.}
    \label{figbsM}
\end{figure}

The entropy density for the black $p$-branes is given by
\begin{eqnarray}\label{eq:densidad-entropia-bs}
    s&=&\frac{S}{V_{p}} =\frac{\pi^2 v r_{+}^2 H_{p}^{p/2}}{2q}\left(\prod_{i=1}^{p}c_{i}^2\right)^{1/2} \ , 
\end{eqnarray}
where $V_{p}$ is the volume of the branes. From here we see that unlike the toroidal black hole solution (which was used as the basis for constructing this solution), the entropy of the black $p$-branes depends on the flavor number due to the presence of the factor $H_{p}$. Fig. \ref{figbsS} shows the behavior of the entropy density of the black string solution in terms of the radius of the event horizon for different values of $N$. Also, from Fig. {\ref{figbsS}} we see that the entropy density is always positive and it increase as we increase the flavor number. 
\\
The mass density of the black $p$-branes is given by 
 \begin{eqnarray}
 \mathrm{m} &=& \frac{M}{V_{p}}=-\frac{\pi  r_{+} v}{16q(p+2)}H_{p}^{p/2}\left(\prod_{i=1}^{p}c_{i}^2 \right)^{1/2}\left(Kq^2 \kappa a_{N}(p+2)+\frac{8}{3}r_{+}^2 \Lambda \right) \ .
\end{eqnarray}
Figure {\ref{figbsM}} shows that, for real solutions (i.e., for solutions whose event horizons satisfy the constraint in Eq. \eqref{rest:m-bs-real}), the mass density is always positive, and for a fixed value of the event horizon it increase as we increase the number of flavors. 
\\
Although, as we have explained before, the study of the perturbative stability of the black $p$-branes is outside the goals of this work, a good approach that can provide clues about the classical stability is the Correlated Stability Conjecture \cite{Harmark:2007md}. This conjecture establishes that the dynamical stability of solutions with extended dimensions, such as the black $p$-branes constructed here, can be related to the local thermodynamic stability, the latter being something that can be analyzed by means of the heat capacity of the solution.

\begin{figure}[h!]
    \centering
    \includegraphics[scale=0.9]{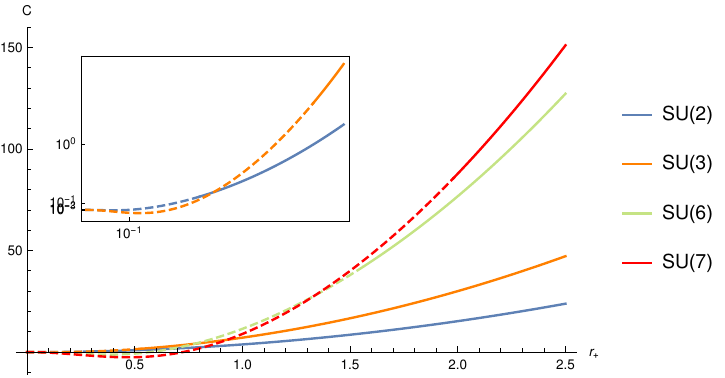} 
    \caption{Heat capacity $C$ of the black string as a function of the event horizon for different values of the flavor number. }
    \label{figbsC}
\end{figure}

\begin{figure}[h!]
    \centering
    \includegraphics[scale=1]{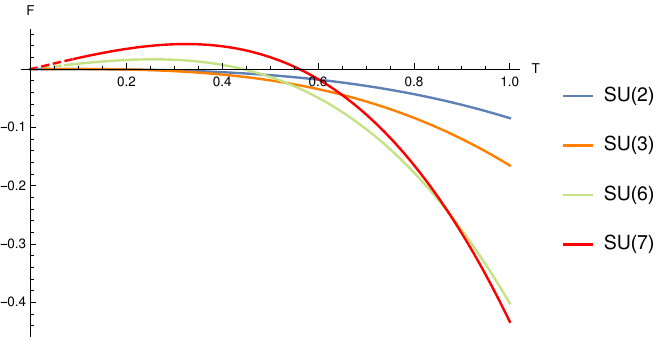} 
    \caption{Free energy $F$ of the black string as a function of the event horizon for different values of the flavor number.}
    \label{figbsF}
\end{figure}

For this purpose we do not need to compactify the extended directions of the black $p$-branes, so we can take the ranges as 
\begin{equation}
     -(2 \pi) L_i \leq z_i \leq (2 \pi) L_i \ , 
\end{equation}
where $L_i \rightarrow \infty$ corresponds to black $p$-branes of infinite extension. 
It follows that the heat capacity is given by
\begin{eqnarray}
   C&=& T\biggl(\frac{\partial S}{\partial T}\biggl) =-\frac{r_{+}^2 \pi^2 v V_{p} H_{p}^{p/2}\left(K\kappa q^2 a_{N}(p+2) +8 r_{+}^2\Lambda \right)}{q\left(K\kappa q^2 a_{N}(p+2) -8 r_{+}^2\Lambda \right)}\left(\prod_{i=1}^{p}c_{i}^2 \right)^{1/2} \ .
\end{eqnarray}
In order to have a positive heat capacity, the radius of the event horizon must satisfy 
\begin{equation}
 r_{+}>\frac{q\sqrt{(p+2)K\kappa a_{N}}}{\sqrt{8(-\Lambda)}}  \ .
\end{equation}
From Fig. \ref{figbsC} we see that the heat capacity for the black string solution is always positive for the values of $r_{+}$ that allow to have a real solution, which indicates the thermodynamic stability of the black string. Based on the Correlated Stability Conjecture the local thermodynamic stability for this solution indicates classical dynamic stability. 
\\
The free energy of the black $p$-branes turns out to be
\begin{equation}\label{F-e-bs}
    \frac{F}{V_{p}}=\mathrm{m}-Ts=-\frac{1}{96}\frac{r_{+}v \pi H_{p}^{p/2}\left( 3K \kappa q^2 a_{N}(p+2)-8r_{+}^2 \Lambda  \right)\left(\prod_{i=1}^{p}c_{i}^2 \right)^{1/2}}{q(p+2)}  \ . 
\end{equation}
It is possible to obtain the free energy as a function of the temperature by inverting Eq. \eqref{T-BS} and substituting it into Eq. \eqref{F-e-bs}. From Fig. \ref{figbsF}, one can see that the free energy has a change of sign that becomes more drastic increasing the number of flavors. For low values of the temperature, the higher value of the free energy is for the configuration with the higher value of $N$. As we increase the temperature a change of sign appears and the lower value of the free energy is also for the higher value of $N$.
\\
\subsection{Self-gravitating instanton}

A regular analytical solution of the Einstein $SU(N)$-Skyrme system can be constructed performing a Wick rotation on the toroidal black hole constructed above.
First of all, it is important to highlight the following fact. In Refs. \cite{Anabalon:2019tcy}, \cite{Banerjee:2007by} and \cite{Cai:2007wz}, using a double Wick rotation of the form 
\begin{equation*}
 t \rightarrow i \phi \ , \qquad \phi \rightarrow i t \ ,  
\end{equation*}
the authors were able to build soliton solutions from a toroidal black hole metric (see also Ref. \cite{Anabalon:2022ksf} for recent applications). One of the key points in such formalism is that the matter field considered (in \cite{Banerjee:2007by}, \cite{Anabalon:2022ksf}, the Maxwell potential) does not depend explicitly on the angular coordinate on which the Wick rotation is being performed. This is different from the case we are considering here, as can be seen from Eq. \eqref{matter1}, where the pionic field depends explicitly on the two angular coordinates. In fact, with the usual double Wick rotation the matter field in Eq. \eqref{matter1} ceases to be an element of $SU(N)$.\footnote{In order to obtain a regular Lorentzian solution one can try to perform an additional "Wick rotation on the hair parameters" , $p \rightarrow i p$, so that $U$ remains an element of $SU(N)$. However, the field equations lead to a constraint between the parameters that cannot be satisfied.}

However, a similar construction can be performed to obtain a regular Euclidean solution of the field equations, that is, a self-gravitating instanton. In fact, let us consider a Wick-like rotation of the form 
\begin{equation} \label{newWick}
  \phi \rightarrow t \ , \qquad t \rightarrow i \phi \ .
\end{equation}
Replacing Eq. \eqref{newWick} into Eqs. \eqref{matter1} and \eqref{g1}, the matter field and the metric become, respectively, 
\begin{equation}
 F_1(x^\mu) = 0 \ , \qquad F_2(x^\mu) =  q \theta \ , \qquad F_3(x^\mu) = p t \ ,   
\end{equation}
\begin{equation}  \label{g3}
 ds^2 =  c^2 r^2 dt^2+ \frac{1}{f(r)} dr^2 + r^2 d\theta^2 + f(r) d\phi^2 \ .
\end{equation}
Then, the field equations are solved by
\begin{equation} \label{f3}
 f(r)= -\frac{K \kappa q^{2} a_{N}}{4} +\frac{m_{0}}{r}  +\frac{K \kappa q^{4} \lambda a_{N}}{32 r^2} - \frac{\Lambda}{3} r^2 \ , 
\end{equation}
where $m$ is an integration constant which is related to the mass of the instanton.
To avoid the conical singularity in the plane $(r,\phi)$, we impose the following periodicity for the angular coordinate $\phi$:
\begin{equation}
    \eta=-\frac{128 \pi r_{0}^{3}}{K \kappa q^{2} a_{N}\left(8 r_{0}+q^{2}\lambda\right)+32 r_{0}^{4}\Lambda}\,,
\end{equation}
where $r_{0}$ is the radius of the instanton satisfying $f(r_{0})=0$. The above guarantees that the solution in Eqs. \eqref{g3} and \eqref{f3} is regular everywhere, and therefore describes a self-gravitating instanton. 
Gravitational instantons are very interesting solutions because these are used as ground states for relevant solutions and could yield new insights into the nature of quantum gravity. 
 
The fact that this solution has been obtained from a toroidal black hole through a rotation of the form in Eq. \eqref{newWick} makes it different from other instanton-type solutions present in the literature \cite{Marino}. Therefore, it is expected that the usual methods for computing the thermodynamics may not apply in this case, and a more exhaustive analysis must be carried out.
From the reason presented above, both the calculation of the thermodynamics and the study of the relevance of this solution in the context of quantum field theory will be explored in a forthcoming paper.
 

\section{Conclusions} \label{sec-5}


In this paper we have constructed a family of new exact solutions of the Skyrme model coupled to Einstein gravity for an arbitrary number of flavors $N$ on the matter field. For this purpose, we have considered the so-called maximal embedding of $SU(2)$ into $SU(N)$ in the Euler angles
parametrization. First, we have constructed analytical solutions describing toroidal black holes with Skyrme hair and arbitrary flavor number $N$. This solution is asymptotically locally anti-de Sitter and generalizes the flat black hole constructed in Ref. \cite{toroidal2} to the case of having non-vanishing Skyrme hair. We have performed the thermal analysis of the black hole, and we have provided with the thermodynamics quantities using the Euclidean approach. We have shown that both the thermodynamics and the geometry strongly depend on the number of flavors of the theory. For a large fixed radius of the event horizon, the most massive configurations are those with the smallest number of flavors. Regarding the entropy $S$, it does not depend explicitly on $N$. For the temperature, we showed that for a fixed value of $r_{+}$, we have that $T$ decreases as we increase $N$. From the plot $F(T)$ for different values of $N$ we saw that as we increase the temperature, transitions start to happen for $N\leq5$. For $N\geq6$ there is no change of sign in the free energy, independent of the value of the temperature. Unlike the case $\lambda=0$ studied in Ref. \cite{toroidal2}, in the present case we showed that the presence of the Skyrme term allows to have transitions, at least for some values of $N$. Indeed, for the flat black hole constructed in \cite{toroidal2}, there are no transitions and the favored configuration is always the one with the higher flavor number.

The second solution that we present is an homogeneous black $p$-brane solution that is constructed by performing a dimensional extension of the toroidal black hole reported in the third section. In order to construct this higher dimensional solution, we need to consider fields that depend linearly on the coordinates of the extended directions. Also, from the field equations we see that the Skyrme coupling must vanish, so, the black hole in the transverse section of the string is the flat black hole solution of Einstein-NLSM presented on \cite{toroidal3}. Also, we see that in order to have a real solution, the radius of the event horizon must have a minimum value, and this restriction makes the mass, entropy and temperature of the solution always positive. We also see that all of these quantities depend on the flavor number. In particular, the mass and entropy density of the solution increase as we increase the flavor number, but for the temperature of the solution, the behavior is the opposite.

Finally, performing a Wick-like rotation on the black hole solution we have constructed a self-gravitating instanton solution. Relevant properties of this solution will be explored in a forthcoming paper.

\section*{Acknowledgments}

The authors are grateful to Fabrizio Canfora and Constanza Quijada for many enlightening comments.
A. V. is funded by FONDECYT postdoctoral Grant No. 3200884. This work was partially funded by the National Agency for Research and Development ANID - SIA Grant No. SA77210097 and FONDECYT Grants No. 11220328 and No. 11220486.  P.C. and E.R. would like to thank to the Dirección de Investigación and Vice-rectoría de Investigación of the Universidad Católica de la Santísima Concepción, Chile, for their constant support. 

\appendix

\section{Regularized Euclidean action}\label{App1}
The regularized Euclidean action is given by the bulk action supplemented with the Gibbons-Hawking term and the counterterms:
\begin{equation}
    I^{\text{E}}=I_{\text{bulk}}+I_{\text{GH}}+I_{\text{ct}} \ , 
\end{equation}
where
\begin{equation}
    I_{\text{bulk}}=-\int d^{4}x\,\sqrt{g}\left( \frac{R-2\Lambda }{2\kappa }+
\frac{K}{4}\mathrm{Tr}[L^{\mu }L_{\mu }]+\frac{K\lambda }{32}\mathrm{Tr}
\left( G_{\mu \nu }G^{\mu \nu }\right) \right)\,,
\end{equation}
\begin{equation}
    I_{\text{GH}}=-\frac{1}{\kappa}\int_{\partial\mathcal{M}}d^{3}x\,\sqrt{h}K_{\rho}\,, 
\end{equation}
\begin{equation}
    I_{\text{ct}}=\frac{1}{\kappa}\int_{\partial\mathcal{M}
}d^{3}x\,\sqrt{h}\left(\frac{2}{ \ell}+\frac{\ell}{2}\mathcal{R}\right)+\frac{K}{4}\int_{\partial\mathcal{M}
    }d^{3}x\,\sqrt{h}\, \ell\, \mathrm{Tr}[L^{i }L_{i}]+\frac{K\lambda}{32}\int_{\partial\mathcal{M}
    }d^{3}x\,\sqrt{h}\, \ell\, \mathrm{Tr}\left(G_{i j}G^{i j}\right)\,.
\end{equation}Here, $h_{ij}$ is the metric induced on the boundary $\partial\mathcal{M}$ at the cut-off $r=\rho$, $K_{\rho}$ is the trace of the extrinsic curvature of the boundary as embedded in $\mathcal{M}$ and $\ell^{2}=-{3/}{\Lambda}$. Additionally to the well-known anti-de Sitter gravitational surface terms in $I_{\text{ct}}$, we add an appropriate generalized counterterm action according to the matter content present in the theory. Taking the limit $\rho\rightarrow+\infty$, we find that the regularized Euclidean action is
\begin{equation} \label{Euclact}
    I^{\text{E}}=-\frac{\beta v\pi r_{+}}{32q}\left(K\kappa\alpha_{N}q^{2}-\frac{4}{3}\Lambda r_{+}^{2}-\frac{3K\kappa \alpha_{N}\lambda q^{4}}{8 r_{+}^{2}}\right)\,,
\end{equation}
where $r_{+}$ is the largest root of the equation $f\left(r_{+}\right)=0=-g_{tt}$.

\section{Holographic mass}\label{App2}
Considering the Brown-York formalism  \cite{Brown:1992br} supplemented with counterterms, we can obtain the energy of the hairy toroidal black hole. In our case, the regularized stress tensor is given by
\begin{eqnarray}
\tau_{ij}=-\frac{1}{\kappa}\left(K_{ij}-h_{ij}K+\frac{2}{\ell}\right)-h_{ij}\ell\left(\frac{K}{4}\, \mathrm{Tr}[L^{i }L_{i}]+\frac{K\lambda}{32}\, \mathrm{Tr}\left(G_{i j}G^{i j}\right) \right)\,.
\end{eqnarray}
Thus, the stress tensor time components are:
\begin{equation}
    \tau_{tt}=\frac{m}{\ell \kappa \rho}+\order{\rho^{-2}} \ .
\end{equation}
Following Brown and York, it is possible to show that the mass, which is the conserved charge associated with time translation, is given by:
\begin{equation}\label{holomass}
    E=\int d\theta d \phi \sqrt{\sigma}u^{i}\tau_{ij}\xi^{j}\,,
\end{equation}
where $\sqrt{\sigma}=\sqrt{g_{\theta\theta}g_{\phi\phi}}$ , with $\sigma$ being the determinant of the induced metric of the surface $t=constant$,
\begin{equation}
    ds^{2}=\sigma_{ij}dx^{i}dx^{j}\,,
\end{equation}
with the normal vector $u^{i}=f^{-1/2}(\partial_{t})^{i}$. In \eqref{holomass}, $\xi^{i}$ is the timelike Killing vector. Then, from the stress tensor we find that
\begin{equation}
    E=\frac{2\pi^{2}v}{q \kappa}m=-\frac{v\pi r_{+}}{48q}\left(3K\kappa a_{N}q^{2}+4\Lambda r_{+}^{2}-\frac{3K\kappa \alpha_{N}\lambda q^{4}}{8 r_{+}^{2}}\right)\, , 
\end{equation}
where we have used $\kappa=8\pi$.


\begin{thebibliography}{99}

\bibitem{Skyrme1}
T.~H.~R.~Skyrme,
Proc. Roy. Soc. Lond. A \textbf{260}, 127-138 (1961).

\bibitem{Skyrme2}
T.~H.~R.~Skyrme,
Nucl. Phys. \textbf{31}, 556-569 (1962).

\bibitem{Witten}
E.~Witten,
Nucl. Phys. B \textbf{223}, 422-432 (1983).

\bibitem{ANW}
G.~S.~Adkins, C.~R.~Nappi and E.~Witten,
Nucl. Phys. B \textbf{228}, 552 (1983).

\bibitem{Luckock:1986tr}
H.~Luckock and I.~Moss,
Phys. Lett. B \textbf{176}, 341-345 (1986).

\bibitem{Droz:1991cx}
S.~Droz, M.~Heusler and N.~Straumann,
Phys. Lett. B \textbf{268}, 371-376 (1991).

\bibitem{Heusler:1992av}
M.~Heusler, S.~Droz and N.~Straumann,
Phys. Lett. B \textbf{285}, 21-26 (1992).

\bibitem{Manton:1981mp}
N.~S.~Manton,
Phys. Lett. B \textbf{110}, 54-56 (1982).

\bibitem{MantonBook}
N. Manton and P. Sutcliffe, Topological Solitons, (Cambridge University Press, Cambridge, 2007).

\bibitem{OliveBook}
David I. Olive and Peter C. West (Editors), Duality and Supersymmetric Theories, (Cambridge
University Press, Cambridge, 1999).

\bibitem{Bizon}
P.~Bizon and T.~Chmaj,
Phys. Lett. B \textbf{297}, 55-62 (1992).

\bibitem{Sawado}
N.~Sawado, N.~Shiiki, K.~i.~Maeda and T.~Torii,
Gen. Rel. Grav. \textbf{36}, 1361-1371 (2004);
N.~Shiiki and N.~Sawado,
Phys. Rev. D \textbf{71}, 104031 (2005);
N.~Shiiki and N.~Sawado,
Class. Quant. Grav. \textbf{22}, 3561-3574 (2005);
H.~Sato and N.~Sawado,
Phys. Lett. B \textbf{660}, 72-79 (2008).

\bibitem{Kunz}
B.~Kleihaus, J.~Kunz and A.~Sood,
Phys. Lett. B \textbf{352}, 247-253 (1995);
T.~Ioannidou, B.~Kleihaus and J.~Kunz,
Phys. Lett. B \textbf{643}, 213-220 (2006).

\bibitem{BD}
Y.~Brihaye and T.~Delsate,
Mod. Phys. Lett. A \textbf{21}, 2043-2054 (2006).

\bibitem{Brihaye}
Y.~Brihaye, C.~Herdeiro, E.~Radu and D.~H.~Tchrakian,
JHEP \textbf{11}, 037 (2017).

\bibitem{Nelmes}
S.~Nelmes and B.~M.~A.~G.~Piette,
Phys. Rev. D \textbf{84}, 085017 (2011).

\bibitem{Pedro}
P.~D.~Alvarez, F.~Canfora, N.~Dimakis and A.~Paliathanasis,
Phys. Lett. B \textbf{773}, 401-407 (2017);
P.~D.~Alvarez, S.~L.~Cacciatori, F.~Canfora and B.~L.~Cerchiai,
Phys. Rev. D \textbf{101}, no.12, 125011 (2020).

\bibitem{crystals}
F.~Canfora,
Eur. Phys. J. C \textbf{78}, no.11, 929 (2018);
F.~Canfora, S.~H.~Oh and A.~Vera,
Eur. Phys. J. C \textbf{79}, no.6, 485 (2019);
F.~Canfora, M.~Lagos and A.~Vera,
Eur. Phys. J. C \textbf{80}, no.8, 697 (2020).

\bibitem{Diego}
F.~Canfora, D.~Hidalgo, M.~Lagos, E.~Meneses and A.~Vera,
Phys. Rev. D \textbf{106}, no.10, 105016 (2022).

\bibitem{CanforaMaeda}
F.~Canfora and H.~Maeda,
Phys. Rev. D \textbf{87}, no.8, 084049 (2013).

\bibitem{toroidal1}
M.~Astorino, F.~Canfora, A.~Giacomini and M.~Ortaggio,
Phys. Lett. B \textbf{776}, 236-241 (2018).

\bibitem{Eiroa}
F.~Canfora, E.~F.~Eiroa and C.~M.~Sendra,
Eur. Phys. J. C \textbf{78}, no.8, 659 (2018).

\bibitem{Daniel}
D.~Flores-Alfonso and H.~Quevedo,
Class. Quant. Grav. \textbf{36}, 154001 (2019).

\bibitem{accretion}
G.~Abbas, H.~Rehman, M.~Usama and T.~Zhu,
Eur. Phys. J. C \textbf{83}, no.5, 422 (2023).

\bibitem{Eloy1}
E.~Ayon-Beato, F.~Canfora and J.~Zanelli,
Phys. Lett. B \textbf{752}, 201-205 (2016).

\bibitem{Eloy2}
E.~Ay\'on-Beato, F.~Canfora, M.~Lagos, J.~Oliva and A.~Vera,
Eur. Phys. J. C \textbf{80}, no.5, 384 (2020).

\bibitem{gravtube1}
F.~Canfora, A.~Giacomini, M.~Lagos, S.~H.~Oh and A.~Vera,
Eur. Phys. J. C \textbf{81}, no.1, 55 (2021).

\bibitem{gravtube2}
B.~Bonga and G.~Dotti,
Phys. Rev. D \textbf{105}, no.4, 044049 (2022).

\bibitem{toroidal2}
M.~Astorino, F.~Canfora, M.~Lagos and A.~Vera,
Phys. Rev. D \textbf{97}, no.12, 124032 (2018).

\bibitem{toroidal3}
C.~Henr\'\i{}quez-B\'aez, M.~Lagos and A.~Vera,
Phys. Rev. D \textbf{106}, no.6, 064027 (2022).

\bibitem{Giacomini:2019qov}
A.~Giacomini and M.~Ortaggio,
JHEP \textbf{09}, 090 (2019).

\bibitem{Bala1}
A.~P.~Balachandran, A.~Barducci, F.~Lizzi, V.~G.~J.~Rodgers and A.~Stern,
Phys. Rev. Lett. \textbf{52} (1984), 887.

\bibitem{Bala2} 
A.~P.~Balachandran, F.~Lizzi, V.~G.~J.~Rodgers and A.~Stern,
Nucl. Phys. B \textbf{256} (1985), 525-556.

\bibitem{Ioa}
T.~A.~Ioannidou, B.~Piette and W.~J.~Zakrzewski,
J. Math. Phys. \textbf{40}, 6223-6233 (1999).

\bibitem{euler1} S. Bertini, S. L. Cacciatori, B. L. Cerchiai, J. Math.
Phys. (N.Y.) 47, 043510 (2006).

\bibitem{euler2} S. L. Cacciatori, F. Dalla Piazza, and A. Scotti, Trans.
Am. Math. Soc. 369, 4709 (2017).

\bibitem{euler3} T. E. Tilma and G. Sudarshan, J. Geom. Phys. 52, 263 (2004).

\bibitem{Gomberoff}
F.~Canfora, A.~Gomberoff, M.~Lagos and A.~Vera,
Phys. Rev. D \textbf{105}, no.8, 084045 (2022).

\bibitem{euler4}
S.~L.~Cacciatori and A.~Scotti,
Universe \textbf{8}, no.10, 492 (2022).

\bibitem{SU(N)}
S.~L.~Cacciatori, F.~Canfora, M.~Lagos, F.~Muscolino and A.~Vera,
JHEP \textbf{12}, 150 (2021);
S.~L.~Cacciatori, F.~Canfora, M.~Lagos, F.~Muscolino and A.~Vera,
Nucl. Phys. B \textbf{976}, 115693 (2022).

\bibitem{Papantonopoulos:2011zz}
E.~Papantonopoulos,
Lect. Notes Phys. \textbf{828} (2011), 1-438.

\bibitem{Caldarelli:2016nni}
M.~M.~Caldarelli, A.~Christodoulou, I.~Papadimitriou and K.~Skenderis,
JHEP \textbf{04} (2017), 001.

\bibitem{Horowitz:1991cd}
G.~T.~Horowitz and A.~Strominger,
Nucl. Phys. B \textbf{360}, 197-209 (1991).

\bibitem{Son:2007vk}
D.~T.~Son and A.~O.~Starinets,
Ann. Rev. Nucl. Part. Sci. \textbf{57}, 95-118 (2007).

\bibitem{Gibbons:1977mu}
G.~W.~Gibbons and S.~W.~Hawking,
Phys. Rev. D \textbf{15} (1977), 2738-2751.

\bibitem{CisternaOliva}
A.~Cisterna and J.~Oliva,
Class. Quant. Grav. \textbf{35}, no.3, 035012 (2018).

\bibitem{Cisterna:2019scr}
A.~Cisterna, C.~Henr\'\i{}quez-B\'aez and J.~Oliva,
JHEP \textbf{01}, 052 (2020).

\bibitem{Cisterna:2018mww}
A.~Cisterna, S.~Fuenzalida, M.~Lagos and J.~Oliva,
Eur. Phys. J. C \textbf{78}, no.11, 982 (2018).

\bibitem{Cisterna:2020kde}
A.~Cisterna, S.~Fuenzalida and J.~Oliva,
Phys. Rev. D \textbf{101}, no.6, 064055 (2020).

\bibitem{GL1}
R.~Gregory and R.~Laflamme,
Phys. Rev. Lett. \textbf{70}, 2837-2840 (1993).

\bibitem{GL2}
R.~Gregory and R.~Laflamme,
Nucl. Phys. B \textbf{428}, 399-434 (1994).

\bibitem{LehnerPretorius}
L.~Lehner and F.~Pretorius,
Phys. Rev. Lett. \textbf{105}, 101102 (2010).

\bibitem{Suzuki:2015axa}
R.~Suzuki and K.~Tanabe,
JHEP \textbf{10}, 107 (2015).

\bibitem{Emparan:2015gva}
R.~Emparan, R.~Suzuki and K.~Tanabe,
Phys. Rev. Lett. \textbf{115}, no.9, 091102 (2015).

\bibitem{Harmark:2007md}
T.~Harmark, V.~Niarchos and N.~A.~Obers,
Class. Quant. Grav. \textbf{24}, R1-R90 (2007).

\bibitem{Anabalon:2019tcy}
A.~Anabalon, D.~Astefanesei, D.~Choque and J.~D.~Edelstein,
JHEP \textbf{07}, 129 (2020).

\bibitem{Banerjee:2007by}
N.~Banerjee and S.~Dutta,
JHEP \textbf{07} (2007), 047.

\bibitem{Cai:2007wz}
R.~G.~Cai, S.~P.~Kim and B.~Wang,
Phys. Rev. D \textbf{76} (2007), 024011.

\bibitem{Anabalon:2022ksf}
A.~Anabal\'on, P.~Concha, J.~Oliva, C.~Quijada and E.~Rodr\'\i{}guez,
Phys. Lett. B \textbf{835}, 137521 (2022).

\bibitem{Marino}
M. Mariño,  Instantons and Large N: An Introduction to Non-Perturbative Methods in Quantum Field Theory,
Cambridge University Press, 2015.

\bibitem{Brown:1992br}
J.~D.~Brown and J.~W.~York, Jr.,
Phys. Rev. D \textbf{47} (1993), 1407-1419.



\end{thebibliography}
\end{document}